\documentclass{ifacconf}

\usepackage{siunitx}
\usepackage{amsmath,bm}
\usepackage{amsfonts}
\usepackage{mathrsfs}
\usepackage{multirow}
\usepackage{booktabs}

\usepackage{tikz}
\usetikzlibrary{matrix}
\usetikzlibrary{shapes.geometric, arrows}

\usepackage{graphicx}      % include this line if your document contains figures
\usepackage{natbib}        % required for bibliography
\begin{document}
\begin{frontmatter}

\title{Bayesian hierarchical modelling for battery lifetime early prediction} 

% \title{Bayesian hierarchical modelling for battery lifetime early prediction\thanksref{footnoteinfo}}

% Title, preferably not more than 10 words.

% \thanks[footnoteinfo]{This work was supported in part by the Chinese Scholarship Council and the Engineering Science Department at the University of Oxford. Thanks also to Masaki Adachi and Dr Nicola Courtier for helpful discussions.}

\author[Second]{Zihao Zhou} 
\author[Second]{David A.\ Howey} 
% \author[Third]{Third C. Author}

%\address[First]{University of Oxford, Parks Road, Oxford OX1 3PJ, UK (e-mail: zihao.zhou@eng.ox.ac.uk).}
\address[Second]{University of Oxford, Parks Road, Oxford OX1 3PJ, UK (e-mail: david.howey@eng.ox.ac.uk).}
% \address[Third]{Electrical Engineering Department, 
%   Seoul National University, Seoul, Korea, (e-mail: author@snu.ac.kr)}

\begin{abstract}                % Abstract of not more than 250 words.
Accurate prediction of battery health is essential for real-world system management and lab-based experiment design. However, building a life-prediction model from different cycling conditions is still a challenge.  Large lifetime variability results from both cycling conditions and initial manufacturing variability, and this---along with the limited experimental resources usually available for each cycling condition---makes data-driven lifetime prediction challenging. Here, a hierarchical Bayesian linear model is proposed for battery life prediction, combining both individual cell features (reflecting manufacturing variability) with population-wide features (reflecting the impact of cycling conditions on the population average). The individual features were collected from the first 100 cycles of data, which is around 5-10\% of lifetime. The model is able to predict end of life with a root mean square error of 3.2 days and mean absolute percentage error of 8.6\%, measured through 5-fold cross-validation, overperforming the baseline (non-hierarchical) model by around 12-13\%.

%The proposed HBM considers both individual cell features and cycling condition features in a two-level structure, which enables it to address large usage variability. The features used were collected from the first 100 cycles of cycling data, which is only around 5-10\% of average lifetime. The results show HBM gives a 3.2 days root mean square error and 8.6\% mean absolute percentage error in an 5-folds cross validation setting, which overperforms the baseline model by around 12\% and 13\% respectively.

% The dataset considered consists of 118 cells from 33 different cycling conditions. HBM gives a 3.1 week median root mean square error in an ordinary in-sample prediction setting and 74\% accuracy in a more challenging out-of-sample classification setting, which overperforms the baseline model by around 20\% and 22\% respectively. 
\end{abstract}


\end{frontmatter}
%===============================================================================

\section{Introduction}
Lithium-ion batteries are ubiquitous due to their relatively long lifetime and high energy density (\cite{cano2018batteries,schmuch2018performance}), but their performance degrades with time and usage. Battery aging behaviours have been widely explored (\cite{birkl2017degradation}), but remain difficult to understand and quantify since batteries are complex electrochemical devices that operate in widely varying conditions. 

% Additionally, collecting test data can be a very time-consuming and complex task, especially when different cycling conditions need to be considered. Except for extreme high C-rate fast aging experiments, it generally takes at least several months to age a battery cell to end of life (EoL), and several cells need to be tested under each selected cycling condition so that the intrinsic cell-to-cell variability may be quantified \cite{dechent2021estimation}.

Existing battery lifetime prediction models can be roughly divided into either physics-based (\cite{reniers2019review}) or data-driven approaches (\cite{sulzer2021challenge}). Physics-based models are built from first principles, accounting for diverse aging mechanisms such as growth of the solid-electrolyte interphase (\cite{liu2020review}), lithium plating (\cite{liu2016understanding}), and particle cracking (\cite{ai2019electrochemical}). While these are good at explaining underlying aging mechanisms, developing models that predict behaviour under different cycling conditions is challenging, since the influential degradation mechanisms may be vary across cycling conditions (\cite{su2016path,raj2020investigation}). 

% Also, physics-based models normally have many parameters to fit, which may cause identifiability problems \cite{reniers2019review,forman2012genetic}.

Data-driven methods directly map from chosen health indicators (features) to battery lifetime (labels) without requiring underlying domain knowledge. For this purpose, (\cite{severson2019data}) proposed an input feature based on discharging curve differences in early cycles, called $\Delta Q(V)$, giving accurate predictions of lifetime using only early cycling data. 
% However, this feature is only effective when deep discharging data are available under controlled circumstances.
Subsequently, other aging features have been proposed based on charging or discharging curve differences (\cite{paulson2022feature}). However, the effectiveness of these may rely on the specific cycling conditions (\cite{sulzerpromise}). 
% Work by Greenbank et al.\ \cite{greenbank2021automated} uses time ratios spent in different signal (voltage, temperature, current and etc) ranges as input features, which can deal with cells under varying cycling conditions.
Alternatively, instead of generating features from the raw signals (voltage/current/temperature) and then using these as inputs for a model, other works implemented sophisticated machine learning methods to directly build models from raw measurements, such as ensemble learning (\cite{li2019lithium}) and long-short memory networks (\cite{zhang2018long}).

Although data-driven models achieve satisfactory results on their datasets, their ability to generalize is still unproven. Most of these models are built at a population level, i.e.\ all battery cells within a dataset follow the same feature-label mapping function, or to put it another way, the model that is learnt is an \emph{average} model for the whole cell population. As a result, these models often lack the ability to give good performance on \emph{individual} cells going through unseen cycling conditions. Recent work (\cite{dechent2021estimation,strange2022automatic}) also shows that cell-to-cell intrinsic manufacturing variability may not be properly addressed by a single population model. To address this, (\cite{deng2022battery}) clustered cells under similar cycling conditions into subgroups, so that the usage variability could be reduced within each subgroup. Then, separate models were built for each subgroup. However, this separate modelling strategy needs to reach a balance between reducing within-group usage variability (more groups) versus group sample size (fewer groups). In reality, separate subgroup models often suffer from a lack of available data and tend to over-fit because of limited samples.

Different from either a single general population model or separate individual models, there is a reasonable middle ground in the form of hierarchical modelling (also called multi-level modelling). Hierarchical models are statistical models that have parameters at more than one level; they have proven to be a useful tool to deal with naturally structured data. For example, in a two-layer model there are parameters at the lower level which are assumed to be drawn from distributions, and these distributions are themselves parameterised by higher level parameters (\cite{gelman2006data}). Hierarchical models have been widely used in different areas including ecology, psychology, sociology and computer vision (\cite{lake2015human,pedersen2019hierarchical}). %They are proven to be a useful tool when dealing with naturally structured data. 
% However, there are only limited studies in the battery field using hierarchical models. Jiang et al.\ \cite{jiang2021Bayesian} implemented a hierarchical Bayesian model to give a prediction for battery lifetime distribution at the cycling protocol level. This model can be viewed as a varying-intercepts model \cite{abdullah2021varying} without any features, which merely builds on cell lifetime observations. The interesting feature-label (health indicators - lifetime) function is not discussed.

In this work, a hierarchical Bayesian model (HBM) is presented to achieve battery life prediction from early life measurements under varying cycling conditions. 
% Specifically, a linear varying-intercepts and varying-slopes parametric model of life versus time \cite{gelman2006data} is built in a Bayesian way. 
The individual cell feature-label relationships can vary across different cycling conditions according to cycling condition variables. Here we combine the MIT battery ageing dataset (\cite{severson2019data}) with extra samples from (\cite{attia2020closed}) to generate a dataset consisting of 169 cells from 61 different cycling conditions. The lifetimes of these 169 cells range from 5 to 80 days. The individual battery health indicators (features) were calculated from the first 100 cycles of cycling data, which is only around 5-10\% of a cell's average lifetime. We show that the HBM overperforms a baseline ridge regression model by 12\% relative root-mean-square error (RMSE) and 13\% mean-percentage-error (MPE). 
% In addition, the HBM model is able to give predictions for cells from unseen cycling conditions, which is harder to achieve by other population-level models.

\section{Data sources and feature extraction}

%In order to explore the impact of usage variability, different cycling conditions need to be included in the selected dataset. Also, the selected features for battery health should have predictive power across different cycling conditions. In this section, the details about data sources and feature selection are discussed.

\subsection{Data sources}
\begin{figure}
\centering
  \includegraphics[width=8.4cm]{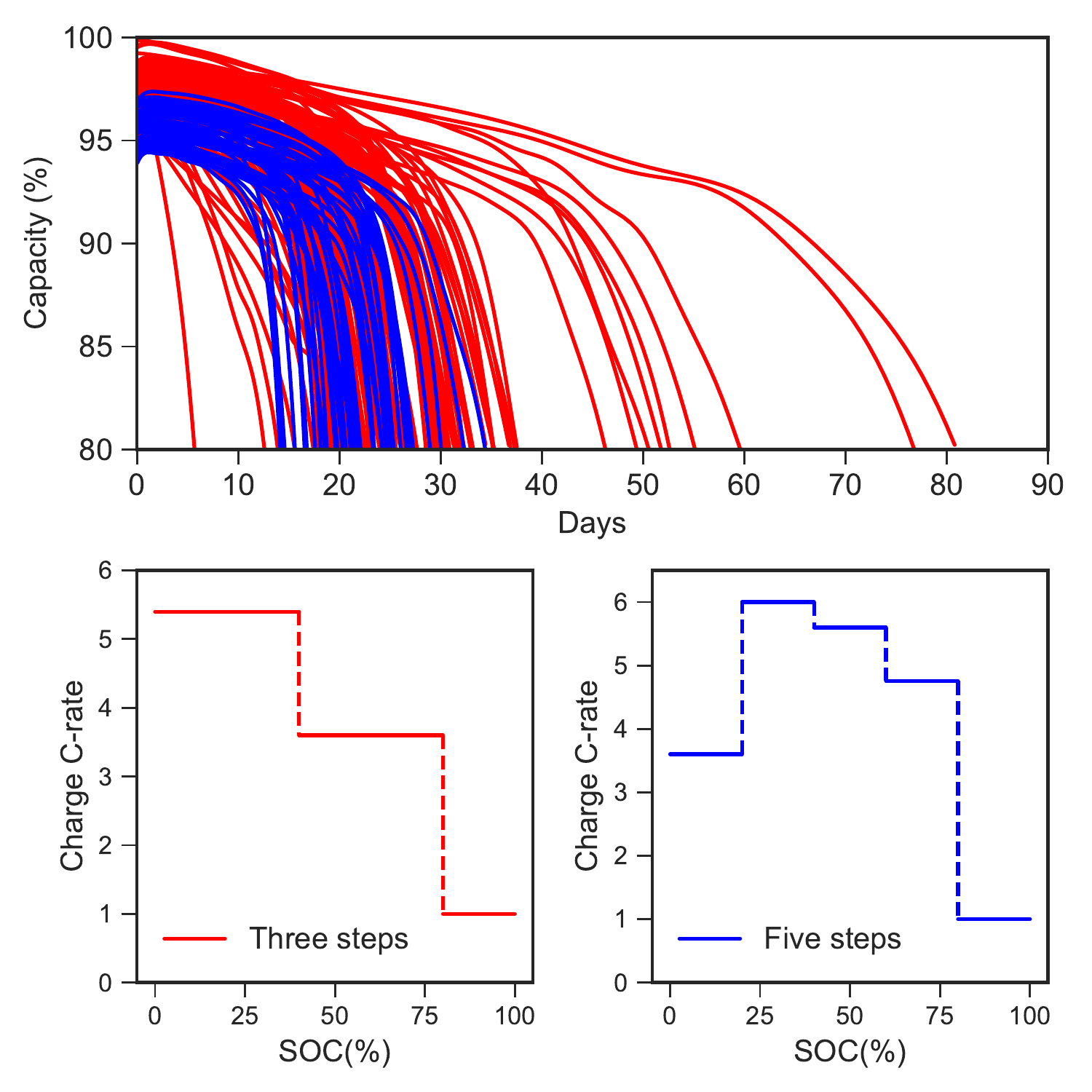}
  \caption{Datasets of (\cite{severson2019data}) (red, three steps) and (\cite{attia2020closed}) (blue, five steps) used here; top shows measured ageing trajectories, bottom shows exemplary individual charging protocols from each dataset respectively;
  in both cases the final step is a 1C constant current charge from 80--100\% SOC.}  
 
  %Top-to-bottom, anti-clockwise: Measured aging trajectories, a typical four steps fast charging protocol from work \cite{severson2019data}, which is denoted as "5.4C-40\%-3.6C" meaning the cell is first charged to 40\% SOC by 5.4C and then changed to 3.6C charging till 80\% SOC. And a six steps fast charging protocol from work \cite{attia2020closed}, which is denoted as "3.6C(20\%)-6C(40\%)-5.6C(60\%)-4.755C(80\%)" meaning the cell is charged with four different C-rates for every 20\% SOC intervals. For both these two protocols, the final two step is 1C constant charging and constant voltage stage.}
  \label{fig:dataset}
\end{figure}

In order to explore the impact of usage variability, different cycling conditions need to be included in the selected dataset. Here, the open source battery cycling datasets of (\cite{severson2019data}) and (\cite{attia2020closed}) are combined together to give a dataset of 169 cells in total. The first dataset (\cite{severson2019data}) consists of 124 lithium iron phosphate/graphite 18650 Li-ion cells (A123) that were cycled at \SI{30}{^\circ C}. The second dataset (\cite{attia2020closed}) is a follow up experiment that contains 45 cells of the same type as before. All these cells underwent identical discharge cycles at 4C but had different fast charging protocols. (In this paper C-rate has the conventional definition, i.e.\ charge or discharge current over nominal capacity.) The cells had a \SI{1.1}{Ah} nominal capacity and the end of life (EoL) is defined as 80\% of nominal capacity remaining.  As shown in Fig~\ref{fig:dataset}, the EoL time covers a wide range from 5 to 80 days (150--2300 cycles), with an average lifetime of 26 days (797 cycles). Cells from the first dataset experienced charging protocols with three steps, while cells from the second set experienced five-step charging protocols. 

\subsection{Feature extraction}
Feature engineering is not the main focus of this work, hence we chose a small number of features at both the group and individual levels in the hierarchical model, guided by existing literature and our prior understanding of battery ageing. In the datasets considered, the charging protocols are quite different from cell to cell. The main difference between them is the sequence of C-rates used during charging, so one way to compare them is to calculate some kind of average charging C-rate for each protocol. Here we used the `SOC-average' charging C-rate as a metric to define differences between protocols---this is related to the total ohmic heat generation ($I^2 R T$, where $R$ is resistance) in the cell during each charging period, normalised by capacity, since
\begin{equation}
\begin{aligned}
&\frac{d(\text{SOC})}{d t}=\frac{I}{Q_0}, \quad d(\text{SOC})=\frac{I}{Q_0} d t \\
\text{hence  } &\int_0^{1} I d(\text{SOC}) = \int_0^{T} I^2 \frac{1}{Q_0} d t,
\end{aligned}
\end{equation}
where the $Q_0$ is the nominal capacity, $T$ the total charge time, and $I$ indicates current. Taking the three steps protocol shown in Fig~\ref{fig:dataset} as an example, the cell is first charged to 40\% SOC at 5.4C, and then charged from 40\% to 80\% SOC at 3.6C. So, the average charging C-rate for this charging protocol is
\begin{equation}
\label{eq:feat_g}
\begin{aligned}
\centering
&q_1=40\%, \quad q_2={40\%}, \quad q_3={20\%}\\
&5.4 \times q_1 + 3.6 \times q_2 + 1 \times q_3  = 3.8\text{C}.
\end{aligned}
\end{equation}
Note that the final 80\% to 100\% SOC stage is always at 1C for all cells within the dataset.

\begin{figure}
\centering
  \includegraphics[width=8.4cm]{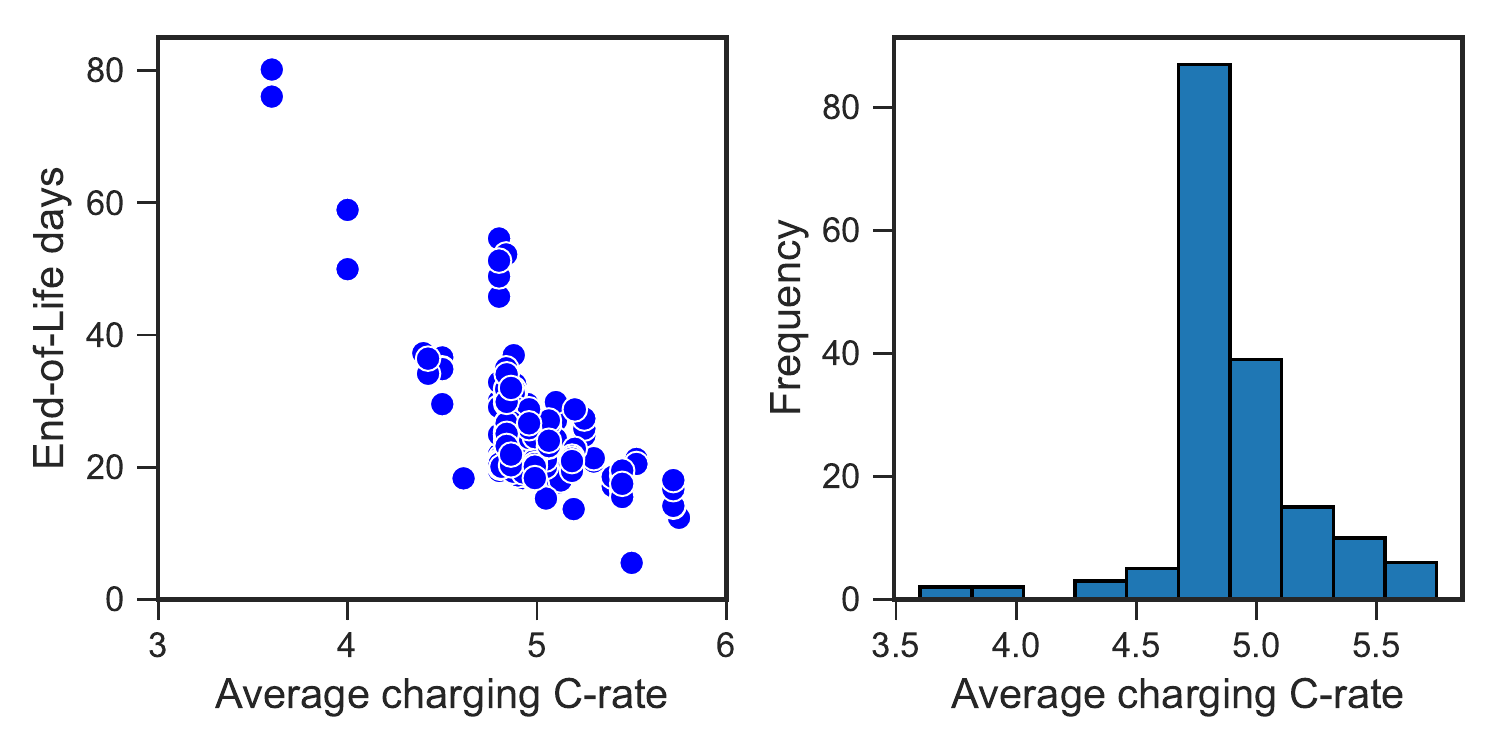}
  \caption{Left shows the relationship between the average charging C-rate and EoL time in days; the absolute linear correlation coefficient is 0.71. Right shows histogram of calculated average charging C-rates.}
  \label{fig:g_feat}
\end{figure}

As shown in Fig~\ref{fig:g_feat}, the average charging C-rate has a strong predictive relationship with the end-of-life (EoL) time in this combined dataset. Multiple cells may have similar or the same average charging C-rate, and these cells will be clustered together into groups later. We denote this average charging C-rate metric as a group level feature $g$, as shown in Table~\ref{Tab:featuredescription}.

\begingroup
\renewcommand{\arraystretch}{1.25}
\begin{table*}
\centering
\caption{Input features used in his work}
\begin{tabular}{lllll}
\hline
\multicolumn{2}{c}{Symbol} & \multicolumn{2}{c}{Description} & \multicolumn{1}{c}{Level} \\ \hline
\multicolumn{2}{l}{$g$} & \multicolumn{2}{l}{Average charging C-rate} & \multicolumn{1}{c}{Group}\\ \hline
\multicolumn{2}{l}{F1} & \multicolumn{2}{l}{Variance of discharge $\Delta Q(V)$ curve between 10th and 100th cycles} & \multicolumn{1}{c}{Individual} \\ \hline
\multicolumn{2}{l}{\begin{tabular}[c]{@{}l@{}}F2 \\ \end{tabular}} & \multicolumn{2}{l}{\begin{tabular}[c]{@{}l@{}} Minimum of discharge $\Delta Q(V)$ curve between 10th and 100th cycles\end{tabular}} & \multicolumn{1}{c}{Individual} \\ \hline
% \multicolumn{2}{l}{\begin{tabular}[c]{@{}l@{}}F3\end{tabular}} & \multicolumn{2}{l}{\begin{tabular}[c]{@{}l@{}}Skewness of discharge $\Delta Q(V)$ curve differences between 10 and 100 cycles\end{tabular}} & \multicolumn{1}{c}{Individual} \\ \hline
% \multicolumn{2}{l}{\begin{tabular}[c]{@{}l@{}}F4\end{tabular}} & \multicolumn{2}{l}{\begin{tabular}[c]{@{}l@{}}Kurtosis of discharge Q-V curve differences between 10 and 100 cycles\end{tabular}} & \multicolumn{1}{c}{Individual} \\ \hline
\multicolumn{2}{l}{F3} & \multicolumn{2}{l}{Discharge capacity at cycle 2} & \multicolumn{1}{c}{Individual} \\ \hline
% \multicolumn{2}{l}{F6} & \multicolumn{2}{l}{Difference between max discharge capacity and cycle 2} & \multicolumn{1}{c}{Individual} \\ \hline
\end{tabular}
\label{Tab:featuredescription}
\end{table*}
\endgroup

\begin{figure}
\centering
  \includegraphics[width=8.4cm]{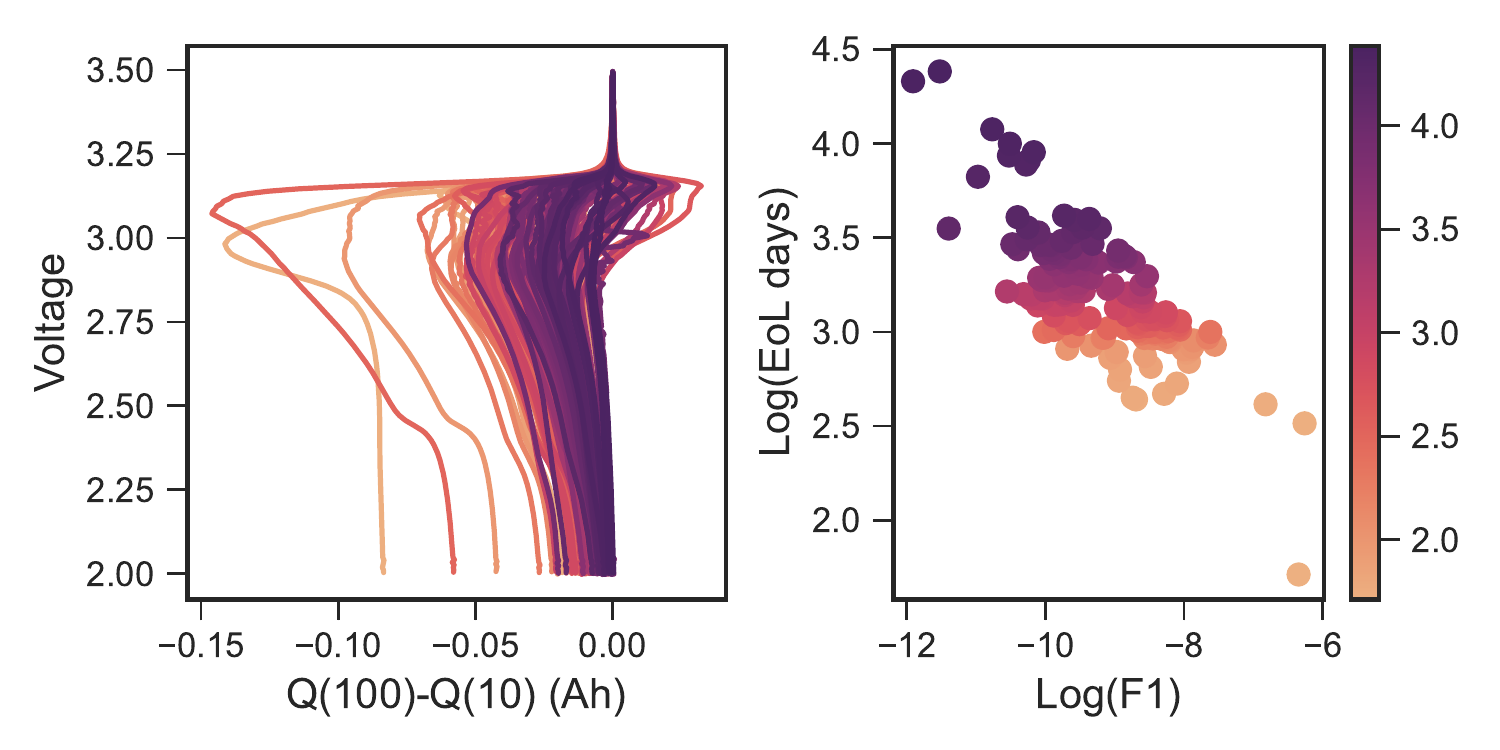}
  \caption{Left shows the feature `F1', the $\Delta Q(V)$ curve between the 10th and 100th discharge cycles; right shows the log-log relationship between the most predictive feature (F1) and the EoL in days. As already reported (\cite{severson2019data}), the correlation is strong, in this case the absolute value of Pearson's linear coefficient is $0.76$.}
  \label{fig:selectedfeatures}
\end{figure}

At the individual cell level, three features which are widely used in existing literature (\cite{severson2019data, paulson2022feature, fei2021early}) were calculated from the first 100 cycles of data for every cell. These are denoted as individual-level features (F1-F3) in Table~\ref{Tab:featuredescription}. The most predictive feature (the one having the largest linear coefficient with log of EoL time), and its relationship with log EoL time, is shown in Fig.\ \ref{fig:selectedfeatures}.
%
% While multiple features can be generated, there may exist severe co-linearity amongst them which is generally considered detrimental for regression tasks. Many methods \cite{hu2020battery,greenbank2021automated, fei2021early} have been implemented to discover relatively uncorrelated features. Principal component analysis (PCA) is often the first choice \cite{murphy2012machine}---it projects the original high-dimensional features into a low-dimensional eigenvector space. The resulting features are linear combinations of the original inputs. 
%
% Here, in order to maintain the interpretability of the original features, an accept-reject strategy based on Pearson's correlation coefficient  \ref{eq:correlation} is adopted as has also been used in a former work \cite{greenbank2021automated}. The correlation coefficient is defined 
% \begin{equation}
% \label{eq:correlation}
% S_{i, j}\left(x_{i}, x_{j}\right)=\left\|\frac{\operatorname{cov}\left(x_{i}, x_{j}\right)}{\sigma\left(x_{i}\right) \sigma\left(x_{j}\right)}\right\|,
% \end{equation}
% where $x_i, x_j$ are individual features, $\operatorname{cov}(x_i,x_j)$ is the covariance between features, and $\sigma(x)$ defines the variance of each individual feature. The detailed process can be summarised as follows: (1) Find the feature correlating best (largest correlation coefficient) with the EoL week; (2) remove all other features which share a correlation coefficient larger than 0.7 with the selected one; (3) repeat the previous step until the required number of features are obtained or there are no more features available. 
%

% \begingroup
% \renewcommand{\arraystretch}{1.25}
% \begin{table}
% \centering
% \begin{tabular}{lll}
% \hline
% Feature name & Feature Description & Feature level \\ \hline
% $x_{\text{feature1}}$ & F1 & Individual cell  \\
% $x_{\text{feature2}}$ & F2 & Individual cell  \\ 
% $g_{\text{chg}}$ & Mean charging C-rate & Group \\ 
% $g_{\text{dis}}$ & Mean discharging C-rate & Group  \\ 
% $g_{\text{DoD}}$ & Mean discharge of depth & Group  \\ \hline
% \end{tabular}
% \caption{Final feature set}
% \label{Tab:finalfeatures}
% \end{table}
% \endgroup

\section{Methodology}

The overall lifetime variability consists of variability caused by usage differences, and variability from manufacturing differences. Ideally there would be many samples for every usage condition so that the cell-to-cell differences under usage could be estimated. However, in the  dataset used, there are only 1-2 cells tested for each different protocol---a sample size too small to give reliable conclusions. To address this, we used constrained K-means clustering to gather cells into usage-related groups each having similar (although not necessarily identical) average charging C-rates. 
%However, many samples have quite similar average charging C-rate, which indicate their similarity in usage conditions. Here, a constrained K-means clustering method is adopted to address this problem. 

\subsection{Cycling conditions clustering}
A commonly used technique to address the balance between individual cell behaviour vs.\ population model accuracy is clustering, i.e.\ grouping of cells into `similar' behavioural subgroups that each contain a relatively large number of cells. However, most previous works (\cite{deng2022battery,jiang2021bayesian}) cluster cells based on features from individual cell voltage signals (or cell EoL lifetime) rather than features representing usage. As a result, cells undergoing quite different cycling conditions may be grouped together. In this work, clustering uses the cycling condition level features ($g$), which enables cells with similar cycling conditions to be grouped together. 

The tool used is constrained K-means clustering (\cite{bhattacharya2018faster}). It is an improved version of K-means that avoids local solutions containing empty clusters, or clusters having very few samples. The goal is to have sufficient sample size (at least 10 cells) in each group, while also maintaining small usage variability within each group (at most 100 cells). % removed:  It fits our needs better than normal K-means, which is sensitive to initialization and may give nearly empty clusters.

\begin{figure}
\centering
  \includegraphics[width=8.4cm]{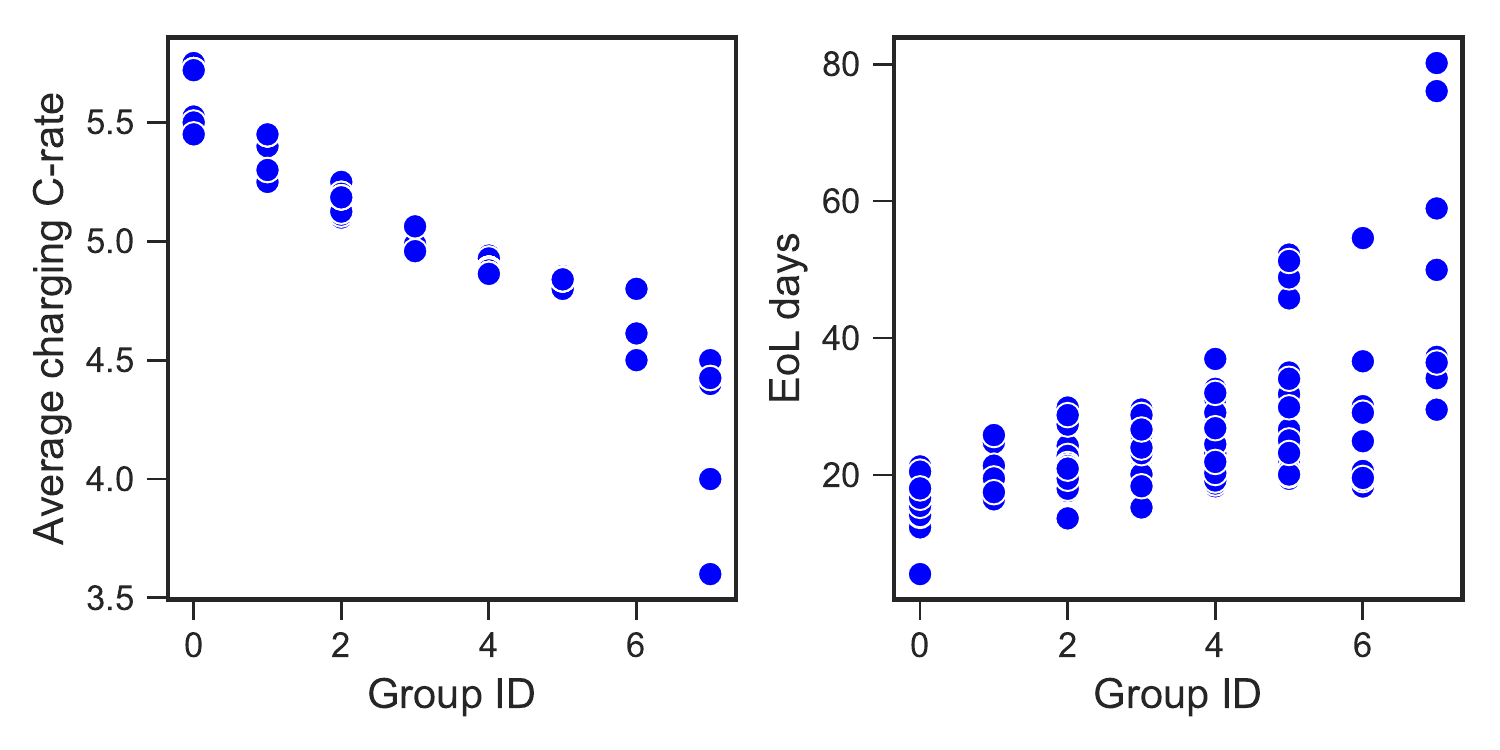}
      \caption{Left shows average charging C-rate for each group with cells clustered into eight groups; right shows associated EoL days for each group. There is relatively large usage variability within the final group due to lack of cells at low ($<$3.5C) charging C-rates.}
    %   In most groups, feature $g$ is concentrated. Because of the lack of cells experience low charging C-rate, the final group has relatively large usage variability.}
  \label{fig:cluster_result}
\end{figure}

There are 169 cells with 61 different average charging C-rates in the dataset, so a reasonable number of clusters is $K \in[4,16]$. Here, $K = 8$ is used as an example (further investigation of the influence of cluster number would be interesting future work). The corresponding clustering result is shown in Fig~\ref{fig:cluster_result}.
% To decide a reasonable cluster number, a practical way called "Elbow" method \cite{kodinariya2013review} is adopted. Here, $K = 5$ is used as an example, as there is a clear changing point for the elbow curve Fig~\ref{fig:elbow} .

\subsection{Bayesian hierarchical linear model}
\begin{figure*}
\centering
    \label{fig:modelstructure}
    \begin{tikzpicture}
        \matrix[matrix of math nodes, column sep=15pt, row sep=40pt] (mat)
        {
            \empty & \empty & \bm{\gamma}& \\ 
            \bm{\theta_{1}},\bm{\sigma_{1}} & \ldots & \bm{\theta_{j}},\bm{\sigma_{j}} & \ldots &\bm{\theta_{J}}, \bm{\sigma_{J}}\\
            y_{1 1}, \ldots, y_{1 i}, \ldots, y_{1 n_{1}} & \ldots & y_{j 1}, 
            \ldots, y_{j i}, \ldots y_{j n_{j}} & \ldots & y_{J 1}, \ldots, y_{J,i}, \ldots y_{J n_{J}}\\
        };

        \foreach \column in {1, 3, 5}
        {
            \draw[->,>=latex] (mat-1-3) -- (mat-2-\column);
            \draw[->,>=latex] (mat-2-\column) -- (mat-3-\column);

        }
        
        \node[anchor=east] at ([xshift =-280pt]mat-1-3) 
        {
        $\begin{aligned}
        &\textbf{Hyper-priors} \\
        &\bm{\gamma} \sim \mathcal{N}(0, 100)\\
    \end{aligned}$};
        
        \node[anchor=east] at ([xshift =-70pt]mat-2-1) 
        {
        $\begin{aligned}
        &\textbf{Cycling Conditions Level-2:} \\
        &\bm{\theta_{j}} \sim \mathcal{N}(\bm{\gamma}^{\top} \bm{g_{j}}, \bm{\sigma_{j}}^2)
    \end{aligned}$};
    
        \node[anchor=east] at ([xshift =-80pt]mat-3-1) 
        {$\begin{aligned}
         &\textbf{Individual Cells Level-1:} \\
         & y_{j i} \sim \mathcal{N}(\bm{\theta_{j}}^{\top} \bm{x_{j i}}, \sigma_{y}^2)
        \end{aligned}$};
    \end{tikzpicture}
     \caption{Hierarchical model structure: parameters can be divided into individual cell level parameters ($\bm{\theta_{j}},\bm{\sigma_{j}}$) and cycling condition level hyper-parameters ($\bm{\gamma}$). While $j$ represents cycling condition group index, $i$ represents individual cell index, $y_{ji}$ represents lifetime of $i$th cell in $j$th cycling group. This two level structure allows the individual cell level feature-label ($x_{ji}-y_{ji}$) relationship to vary across different cycling conditions based on cycling condition level features ($\bm{g_{j}}$).}
     \label{fig:modelstructure}
\end{figure*}
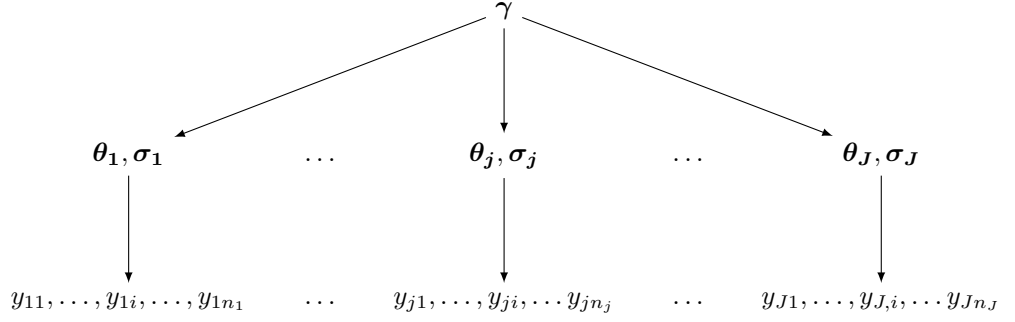
We now discuss the construction of the hierarchical model for life prediction. To motivate the need for a hierarchical model, one should first check whether the group level effect is significant in the target dataset. The variance partition coefficient (VPC), Equation \ref{eq:vpc}, is a metric which represents the percentage of label variance explained by including a grouping effect. 
\begin{equation}
\centering
\label{eq:vpc}
\begin{aligned}
&\text{VPC}=\frac{\sigma_{\text {group}}^{2}}{\sigma_{\text {individual }}^{2}+\sigma_{\text {group}}^{2}} \\
&\sigma_{\text {group}}^{2}=\operatorname{var}\left(\bar{u}_{j}-\bar{u}\right),   \bar{u}=\frac{1}{N} \sum_{j=1}^{k} \sum_{i=1}^{n_{j}} y_{j i}\\
&\sigma_{\text {individual}}^{2}=\operatorname{var}\left(y_{ji}-\bar{u}_{j}\right), \bar{u}_{j}=\frac{1}{n_{j}}\sum_{i=1}^{n_{j}} y_{j i}
\end{aligned}
\end{equation}

Here, $\sigma_{\text{group}}$ defines the variance between groups, and $\sigma_{\text{individual}}$ defines the variance of the whole population where each cell is compared to the mean of its group, $\bar{u}$ is the average lifetime for all cells in the population, $\bar{\mu}_j$ is the average lifetime of the $j$th group, $\operatorname{var}$ has the usual definition of sample variance, $y_{ji}$ is all individual lifetimes for the $i$th cell within the $j$th group, $n_j$ is the total number of cells within the $j$th group and $N$ is the total number of cells in the population. When the group variance $\sigma_{\text {group}}^{2} = 0$, there is no difference between groups, that is, every group has same mean lifetime compared to the overall population battery mean lifetime. When the $\sigma_{\text {individual }}^{2} = 0$, there is no differences within a group, that is, within each group, samples share the same lifetime value. For our dataset, $\text{VPC}=74.5\%$, which denotes that 74.5\% of the lifetime variance is caused by the cycling condition group effect. 

% Given the large influence from cycling condition group, it is reasonable to apply a hierarchical model to our dataset. Hierarchical linear models (HLM) are often employed when dealing with naturally structured data. These models account for not only relationships between features and labels but also relationships between different levels of features \cite{gelman2006data}. The traditional ordinary least square (OLS) regression method can be viewed as a special case of a HLM when the features are themselves independent \cite{woltman2012introduction}. For example, considering students in different classes, one may build a regression model to predict their grades based on learning hours per week. However, the exact relationship between learning hours and grades may vary across different classes because of different teaching styles in in each class.

Given the large influence from cycling condition group, it is reasonable to apply a hierarchical model. The generative model structure is described by the schematic in Fig.\  \ref{fig:modelstructure}. In this framework, within the $j$th group, the first (lower) level inference is done for individual cell lifetime regression parameters ($\bm{\theta_{j}}$) where $\theta_{j0}$ is a scalar representing the mean lifetime within the $j$th group and $\bm{\theta_{j}}$ is a $4\times1$ vector because the number of features is three. In this case because we only have three features, we could have explicitly written the lower level forward model for the $j$th group and $i$th cell within that group as 
\begin{equation}
y_{i j} \sim \mathcal{N}(\theta_{j0} + \theta_{j1} x_{i1}^{j} + \theta_{j2} x_{i2}^{j}  + \theta_{j3} x_{i3}^{j} , \sigma_{y}^2)
\end{equation}
Since the EoL time $y_{ij}$ is accurately measured in the dataset, the sample noise variance $\sigma_{y}$ was fixed to 1 here (it can also be assigned as a distribution and learnt from the dataset).
%assigned a standard exponential distribution $\sigma_{y}\sim \operatorname{Exp}(0,1)$.
For the sake of flexibility and compactness we use vector notion. The second (upper) level inference is undertaken for cycling condition parameters ($\bm{\gamma_0},\bm{\gamma_1},\bm{\gamma_2},\bm{\gamma_3}$), where the $\bm{\gamma_0},\bm{\gamma_1},\bm{\gamma_2},\bm{\gamma_3}$ parameters are each $2\times1$ vectors (because each group has one group level feature and an intercept parameter), defined as 
% Equation \ref{eq:notions} below.
\begin{equation}
\begin{aligned}
\centering
&\bm{\theta_{j}}=\left[\begin{array}{l}
\theta_{j 0} \\
\theta_{j 1} \\
\theta_{j 2} \\
\theta_{j 3}
\end{array}\right] \bm{g_{j}}=\left[\begin{array}{l}
1 \\
\frac{1}{n_{j}}\sum_{i=1}^{n_{j}} g_{j i}
\end{array}\right]  
\bm{x_{j i}}=\left[\begin{array}{c}
1 \\
x_{i 1}^{j} \\
x_{i 2}^{j} \\
x_{i 3}^{j}
\end{array}\right] \\  
&\bm{\gamma_{0}}=\left[\begin{array}{l}
\gamma_{0,0} \\
\gamma_{0,g} \\
\end{array}\right] \bm{\gamma_{1}}=\left[\begin{array}{l}
\gamma_{1,0} \\
\gamma_{1,g} 
\end{array}\right] \bm{\gamma_{2}}=\left[\begin{array}{l}
\gamma_{2,0} \\
\gamma_{2,g}
\end{array}\right] \bm{\gamma_{3}}=\left[\begin{array}{l}
\gamma_{3,0} \\
\gamma_{3,g} 
\end{array}\right]\\
& \bm{\gamma} = \left[
\begin{array}{l}
\bm{\gamma_0},\bm{\gamma_{1}},\bm{\gamma_{2}},\bm{\gamma_{3}}
\end{array}\right], Y_{j}=\left(y_{j 1}, \ldots, y_{j n_{j}}\right) \\
& \left\{Y\right\}=\left\{Y_{1}, \ldots Y_{j}\right\},
\end{aligned}
\label{eq:notions}
\end{equation}
where $Y_j$ is an output vector of lifetimes for all cells within the $j$th group.

The individual slopes of each group model, $\bm{\theta_{j}}$, are assumed to be drawn from normal distributions centered at $\bm{\gamma}^{\top}\bm{g_{j}}$, 
\begin{equation}
\bm{\theta_j} \sim \mathcal{N}\left(\bm{\gamma}^{\top} \bm{g_j}, \bm{\sigma_j}^2\right),
\end{equation}
where $\bm{g_{j}}$ is a $2\times1$ vector of known specific mean features for the $j$th group (i.e., the mean of the individual average charging C-rate for each group). In the top layer, weakly informative (i.e., wide) priors are assigned for all hyper-parameters. %Notice that, $\bm{\sigma_j}$ is assumed to have no influence from cycling conditions ($\bm{g_j}$), which means they are learnt from samples of $j$th group only.
% To summarise, parameters are written in vector form, and observations are written in matrix form, for the $j$th group and $i$th cell within that group, as follows,

Our objective is to infer posterior distributions for both the group models and the population model, $P\left(\bm{\theta_{j}} \mid Y_{j} \right)$ and $P\left(\bm{\gamma} \mid\left\{Y\right\}\right)$ respectively. The overall process is
\begin{enumerate}
\item Calculate level-2 posterior distribution $P\left(\bm{\gamma} \mid\left\{Y\right\}\right)$;
\item Use level-2 posterior distribution as prior for level-1 parameters, calculate level-1 posterior distribution $P\left(\bm{\theta_{j}} \mid \bm{\gamma}, Y_{j}\right)$;
\item Use level-1 posterior distribution to make prediction on individual labels.
\end{enumerate}

Using Bayes' rule, the level-2 posterior can be estimated by multiplying the likelihood function with the hyper-prior:
\begin{equation}
\begin{aligned}
&P\left(\bm{\gamma} \mid\left\{Y\right\}\right) \propto \text { Likelihood } \times \text{Prior}\\
&\text { Likelihood }=\prod_{j=1}^{J} P\left(Y_{j}\mid \bm{\gamma} \right) \\
&\text{Prior} = \mathcal{N}(\mathbf{0}, S_{0}), 
        S_{0} = 100 \cdot I
\end{aligned}
\label{eq:mainposterior}
\end{equation}

Since the level-2 hyperparameters $\bm{\gamma}$ are not directly related to individual observations, the likelihood function can be calculated by integrating out the level-1 parameters, giving %
%
% \begin{equation}
% P\left(Y_j \mid \gamma\right)=\int_{\boldsymbol{\theta}_j} P\left(Y_j \mid \boldsymbol{\theta}_j\right) \cdot P\left(\boldsymbol{\theta}_j \mid \gamma\right) d \boldsymbol{\theta}_j
% \label{eq:integrator}
% \end{equation}
%
\begin{equation}
\begin{aligned}
P\left(Y_j \mid \bm{\gamma}\right) &=\int_{\bm{\theta_j}} P\left(Y_j \mid \bm{\theta_j}\right) \cdot P\left(\bm{\theta_j} \mid \bm{\gamma}\right) d \bm{\theta_j} \\
P\left(Y_j \mid \bm{\theta_j}\right) &=\prod_{i=1}^{n_j} \mathcal{N}\left(\operatorname{err_i}\left(\bm{\theta_j}\right) ; 0,1\right) \\
\operatorname{err_i}\left(\bm{\theta_j}\right) &=\left\{\bm{\theta_j}^{\top} x_{i j}-y_{i j}\right\}.
\end{aligned}
\label{eq:integrator}
\end{equation}

% where
% \begin{equation}
% \begin{aligned}
% P\left(Y_{j} \mid \mu_{j}, \bm{\theta_{j}}\right) &=\prod_{i=1}^{n_{j}} \mathcal{N}\left(\mu_{j} + \bm{\theta_{j}}^{\top} x_{j i}, \sigma_{y}^{2}\right) \\
% &=\mathcal{N}\left(\mu_{j} + \bm{\theta_{j}}^{\top} \bm{\bar{x}_{j}},  \frac{\sigma_{y}^{2}}{n_{j}}\right),
% \end{aligned}
% \end{equation} %
%
% where we exploit the fact that $n_{j}$ Gaussian measurements with mean value $\mu_{j} + \bm{\theta_{j}}^{\top} x_{j i}$ and variance $\sigma_{y}^{2}$ are equivalent a Gaussian with mean $\bm{\theta_{j}}^{\top} \bm{\bar{x}_{j}}$ and variance $\frac{\sigma_{y}^{2}}{n_{j}}$. 
Notice that the level-2 likelihood component $P\left(Y_{j} \mid \bm{\gamma}\right)$ is the model evidence (marginal likelihood) for the level-1 model of $j$th subgroup, which indicates there is an averaging effect for level-1 model selection of each sub-group in our proposed two level structure. 

% Equation \eqref{eq:integrator} can be analytically calculated by completing the square and using the Woodbury lemma \cite{woodbury1950inverting}. The solution is given by

% \begin{equation}
% \begin{aligned}
% &P\left(Y_{j} \mid \mu, \bm{\gamma}\ \right)=\mathcal{N}\left(M_{j}, \Sigma_{j}\right) \\
% &\begin{split}
%     M_{j}&=n_{j} \sigma_{y}^{-2}\bm{\bar{x}_{j}}^{\top}\left[\begin{array}{c}
% \sigma_{0}^{-2} u \\
% \bm{\sigma_{\theta}}^{-2} \cdot \bm{\gamma^{\top}} \bm{g_{j}}
% \end{array}\right]\\
%     &\left(\sigma_{y}^{-2} \bm{\bar{x}_{j}}\bm{\bar{x}_{j}}^{\top}+\left[\begin{array}{cc}
%     \sigma_{0}^{-2} & 0 \\
%     0 & \bm{\sigma_{\theta}}^{-2}
%     \end{array}\right]\right) \Sigma_{j}^{-1}
% \end{split} \\
% &\Sigma_{j}=\frac{\sigma_{y}^2}{n_{j}} +\bm{\bar{x}_{j}}^{\top} \left[\begin{array}{cc}
%     \sigma_{0}^{-2} & 0 \\
%     0 & \bm{\sigma_{\theta}}^{-2}
%     \end{array}\right] \bm{\bar{x}_{j}}
% \end{aligned}
% \label{eq:analyticlikelihood}
% \end{equation}

Given the level-2 likelihood expression (\ref{eq:integrator}), the posterior distribution for hyper-parameters $P\left(\bm{\gamma} \mid\left\{Y\right\}\right)$ can be estimated by Markov chain Monte Carlo (MCMC). After that, the posterior distributions for level-1 parameters $P\left(\bm{\theta_{j}} \mid \bm{\gamma}, Y_{j}\right)$ can be estimated by applying Bayes' rule, 
\begin{equation}
P\left(\bm{\theta_j} \mid \bm{\gamma}, Y_j\right)=\frac{P\left(Y_j \mid \bm{\theta_j}\right) \cdot P\left(\bm{\theta_j} \mid \bm{\gamma}\right) \cdot P(\bm{\gamma})}{P\left(Y_j \mid \bm{\gamma}\right)}.
\end{equation}

The whole inference process was performed in Python using the PYMC3 package \cite{salvatier2016probabilistic}.

\subsection{Baseline comparison and evaluation metrics}
The performance of the approach was evaluated by comparing against a baseline ridge regression model under two different error metrics, defined below. To give a richer comparison, the ridge model was built on two different feature sets: one has 6 features (denoted as the ``discharge'' model in \cite{severson2019data}), and the other has 4 features (including $g$ and F1-F3, the same features as the HBM). Notice that the 6 features used in the ``discharge'' model also include F1-F3 used in this paper.

The first error metric is the root mean square error (RMSE) of the EoL predictions. This is a commonly used metric for predictive performance, however, it is significantly influenced by outliers. Therefore, the second metric is mean absolute percentage error (MAPE), which measures the percentage difference in EoL time between the predicted $y_\text{pred}$ and observed end of life $y_\text{real}$. The two are defined by
\begin{equation}
\begin{aligned}
&\text { RMSE }=\sqrt{\frac{\sum_{i=1}^{N}\left(y_{\text {true }}^{i}-y_{\text {pred }}^{i}\right)^{2}}{N}}, \\
&\text { MAPE }=\frac{100}{N} \sum_{i=1}^{N} \left|\frac{y_{\text {true }}^{i}-y_{\text {pred }}^{i}}{y_{\text {true }}^{i}}\right|.
\end{aligned}
\label{eq:metrics}
\end{equation}

K-fold cross validation is a commonly used paradigm to evaluate predictive ability (\cite{arlot2010survey}). Here, the train and test sets were selected with 5-fold cross-validation, resulting in a 135 cells training set and a 34 cells test set. 
%Notice that, because of the in-group prediction setting, the 5-fold cross validation is operated for each group, which assumes we know the cycling condition information (average charging C-rate in this case) for the test samples. 
Then, 5-fold cross validation was performed 4 times (totalling 20 RMSE values) to observe the general predictive ability of the proposed model.

\begin{figure}
    \centering
    \includegraphics[width=8.4cm]{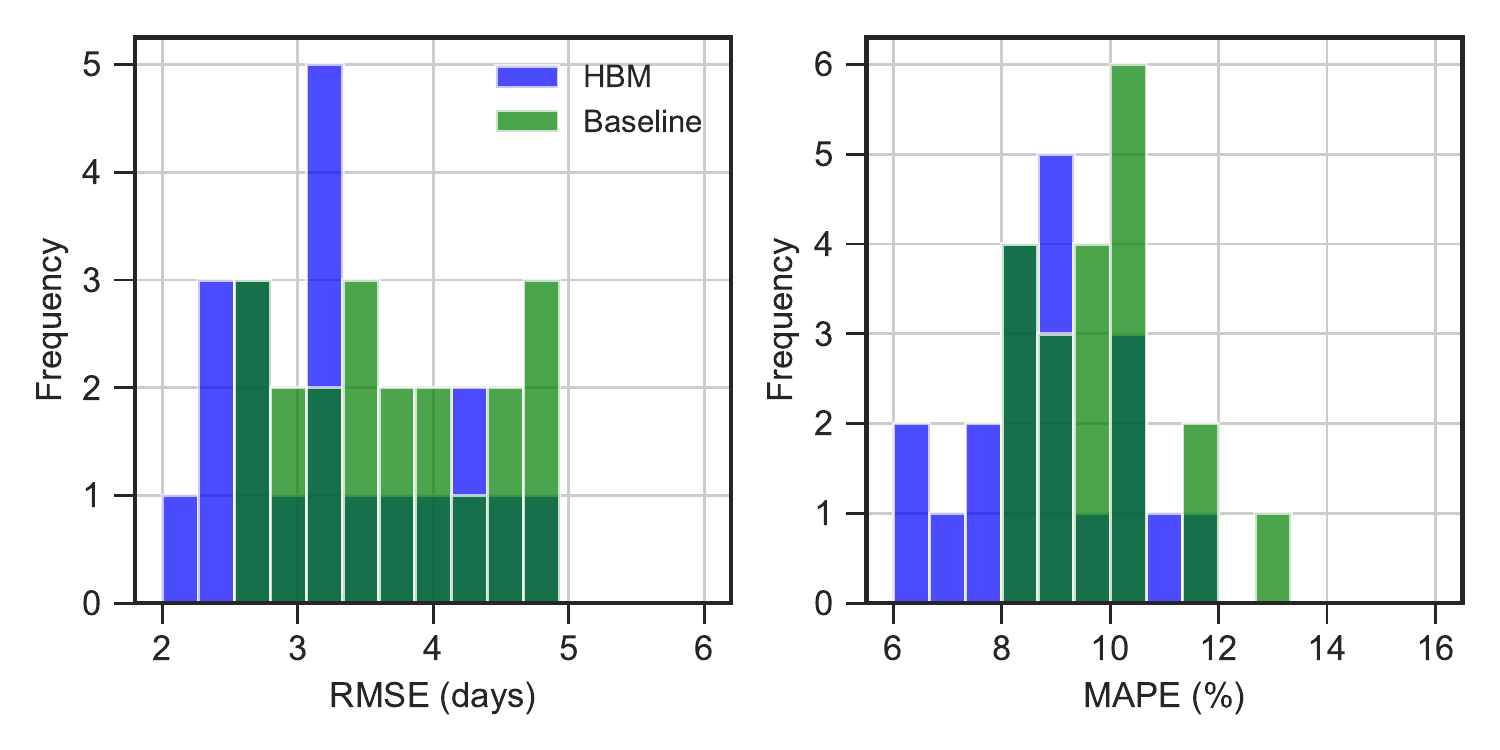}
    \caption{Histograms of all results, as RMSE and MAPE}
    \label{fig:summaryhistogram}
\end{figure}

Table \ref{Tab:results} summarises the results. The HBM gives 3.20 days median RMSE and 8.6\%  median MAPE, surpassing the baseline model on both feature sets. Notice that, by including usage feature $g$, the performance of the baseline model is also improved, which indicates the importance of cycling condition as a feature. More visualization of the results is shown in Fig.\ \ref{fig:summaryhistogram}. For the HBM, the majority (70\%) of the RMSE values on EoL time are below 3.5 days. For the baseline model, 50\% of trials give RMSE values larger than 3.5 weeks. The case for MAPE is also similar. The HBM generally gives around 12-13\% improvement in performance compared to the original baseline model in both RMSE and MAPE.

\section{Results and Discussion}
\begin{table*}[ht]
\centering
\caption{Summary of model predictive performance (RMSE is in days)}
\begin{tabular}{cccccc}
\hline
\multicolumn{2}{l}{} & Baseline (original) & Baseline (with feature $g$) & HBM & Improvement\\ \hline
\multicolumn{1}{l}{\multirow{2}{*}{RMSE}} & Median & 3.67 & 3.38 & 3.20 & 12.8\%  \\ 
\multicolumn{1}{l}{} & Mean & 3.72 & 3.49 & 3.26 & 12.4\% \\ \hline
\multicolumn{1}{l}{\multirow{2}{*}{MPE}} & Median & 9.9\% & 9.1\% & 8.6\% & 13.1\% \\ 
\multicolumn{1}{l}{} & Mean & 9.9\% & 9.2\% & 8.7\% & 12.1\% \\ \hline
\end{tabular}
\label{Tab:results}
\end{table*}

\begin{figure}
\centering
  \includegraphics[width=8.4cm]{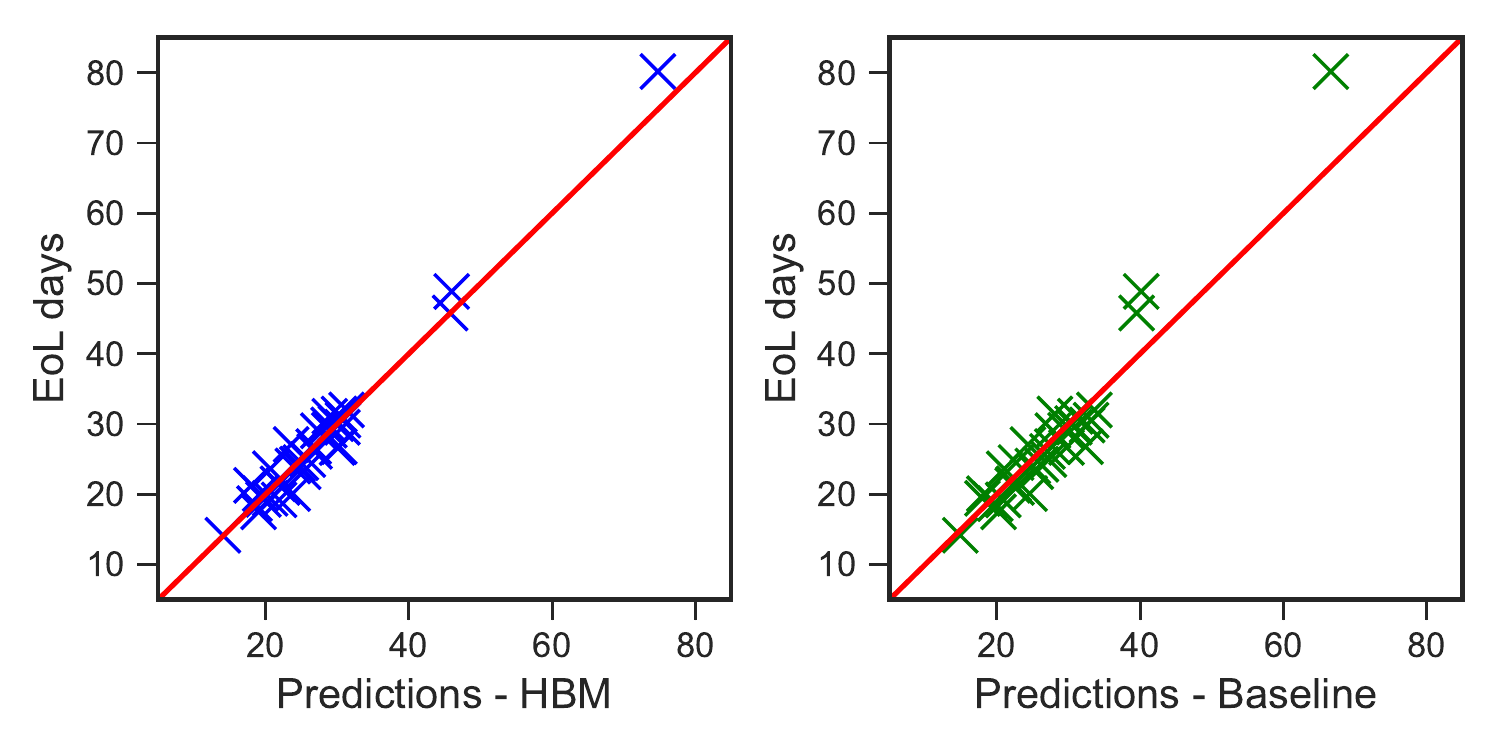}
  \caption{Scatter plot of predicted versus actual EoL (days) for HBM vs.\ baseline model}
  \label{fig:predictions}
\end{figure}

Fig.~\ref{fig:predictions} shows a scatter plot of predictions versus actual EoL time from one trial for the HBM and the baseline model. Both give relatively good predictions in the mid-life range (around 30 days). However, the baseline model fails to give good predictions for long-life ($>$40 days) cells. The HBM has a reasonably consistent predictive performance across the whole lifetime range.

\begin{figure}
\centering
  \includegraphics[width=8.4cm]{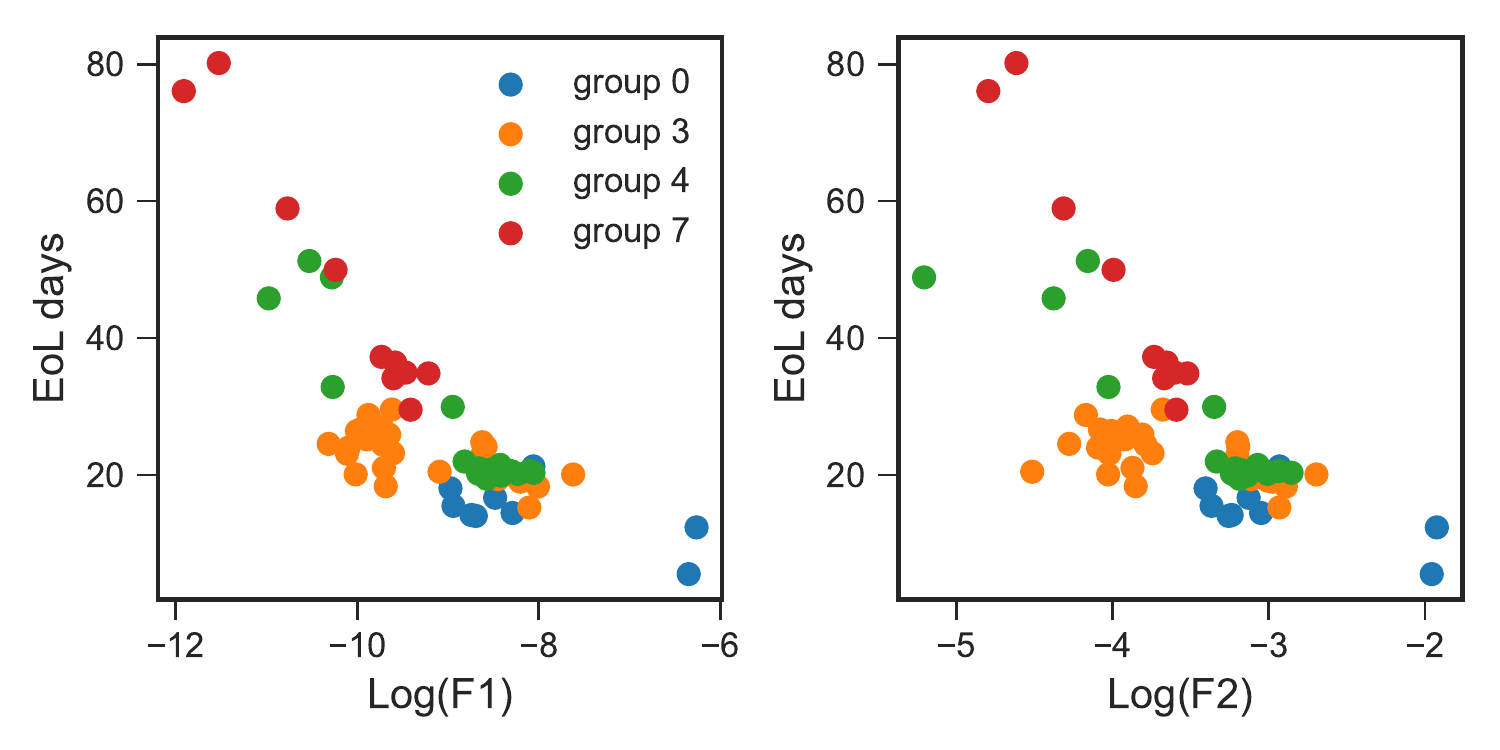}
  \caption{Relationships between EoL vs.\ individual cell features and group labels based on different cycling conditions. Groups $0, 3, 4, 7$ are selected to show the influence of cycling conditions. See also Fig~\ref{fig:cluster_result}.}
  \label{fig:clusterrelationship}
\end{figure}

The superior performance of the proposed HBM can be explained by its ability to account for relationships between different levels of features, or in other words, the inherent structure within the dataset. The cycling condition level feature influences the relationships between individual cell level features. Fig.\ \ref{fig:clusterrelationship} shows cells from four different cycling condition clusters---while all of them share negative correlations between features and labels, their slopes and input feature ranges are quite different. As the average charging C-rate decreases from group $0$ to group $7$, the slope become steeper and the range of log F1 values  decreases.

In a conventional regression setting, the cycling condition features and individual cell features are treated in the same way, assuming they are independent. As a result, the inherent relationships between these two different categories of features are ignored. In contrast, in hierarchical approaches, this kind of inherent relationship is explicitly modelled, which enables better model performance when facing structured data.

\section{Conclusion }
A hierarchical Bayesian linear model was proposed to address the problem of battery early life prediction under varying usage conditions and using only the first 100 cycles data (only 5-10\% of the entire life) as inputs. The proposed HBM was tested on a open data set consisting of 169 cells experiencing 61 different cycling protocols. In a 5-fold cross-validation experiment, the HBM gives a 3.20 days median RMSE and a 8.6\% median MAPE, which overperforms the baseline model by 12\%. 
% In practical out-of-sample classification setting, HBM gives 74\% accuracy which over-performs the baseline model by over 20\%. 
These results show the effectiveness and further potential of the hierarchical models for battery early life prediction.

There are two directions for future work. First, the influence of cluster number and cluster size needs to be further investigated, as these two parameters are known to be important in many other fields when using hierarchical models (\cite{gelman2006data}). In our dataset, eight clusters is a reasonable number to balance within-group usage variability versus sample size. However, this was an empirical choice based mainly on trial and error. Second, more different usage conditions need to be included. In the used dataset, the usage differences only locate in charging period. The influence from discharge C-rate, depth of discharge and rest periods are not included. More useful group level features may be found by including more usage conditions. Third, nonlinear functions, or even non-parametric models such as Gaussian processes, could be implemented hierarchically, and may give better performance compared to linear regression models. A further collaboration work with Prof.\ Hu Chao's group aims to address above challenges is under preparation.

\begin{ack}
This work was supported in part by the Chinese Scholarship Council and the Engineering Science Department at the University of Oxford. Thanks to Tingkai Li, Adam Thelen and Prof.\ Hu Chao for the helpful discussion about model details. Thanks to Dr Nicola Courtier for suggestions about feature extraction. Thanks also to Masaki Adachi for proofreading.
\end{ack}

\bibliography{ifacconf}             % bib file to produce the bibliography
                                                     % with bibtex (preferred)
                                                   
%\begin{thebibliography}{xx}  % you can also add the bibliography by hand

%\bibitem[Able(1956)]{Abl:56}
%B.C. Able.
%\newblock Nucleic acid content of microscope.
%\newblock \emph{Nature}, 135:\penalty0 7--9, 1956.

%\bibitem[Able et~al.(1954)Able, Tagg, and Rush]{AbTaRu:54}
%B.C. Able, R.A. Tagg, and M.~Rush.
%\newblock Enzyme-catalyzed cellular transanimations.
%\newblock In A.F. Round, editor, \emph{Advances in Enzymology}, volume~2, pages
%  125--247. Academic Press, New York, 3rd edition, 1954.

%\bibitem[Keohane(1958)]{Keo:58}
%R.~Keohane.
%\newblock \emph{Power and Interdependence: World Politics in Transitions}.
%\newblock Little, Brown \& Co., Boston, 1958.

%\bibitem[Powers(1985)]{Pow:85}
%T.~Powers.
%\newblock Is there a way out?
%\newblock \emph{Harpers}, pages 35--47, June 1985.

%\bibitem[Soukhanov(1992)]{Heritage:92}
%A.~H. Soukhanov, editor.
%\newblock \emph{{The American Heritage. Dictionary of the American Language}}.
%\newblock Houghton Mifflin Company, 1992.

%\end{thebibliography}

\end{document}